\documentclass{article}
\usepackage{graphicx}
\usepackage[hang]{caption}
\usepackage{subcaption} 
\usepackage[utf8]{inputenc}
\usepackage[T1]{fontenc}
\usepackage[english]{babel}
\usepackage{microtype} % optional, for aesthetics
\usepackage{tabularx} % nice to have
\usepackage{booktabs} % necessary for style
\usepackage{graphicx}
\usepackage{indentfirst}
\usepackage{amssymb}
\usepackage{amsmath}

\usepackage{bm}
\usepackage{braket}
\usepackage[margin=1cm]{caption}

\mathchardef\normalvee=\vee

\renewcommand{\vee}{{\mathpalette\scaledvee\relax}}
\newcommand{\scaledvee}[2]{\scalebox{2}[1]{$#1\normalvee$}}

\newcommand{\W}{\mathtt{\lor{}\hspace{-1.9pt}\/\lor}}%{\larger${{\mathsf{w}}}$\smaller}%{$\lor\hspace{-2pt}\lor$}
\newcommand{\VW}{\mathtt{\lor{}\hspace{-3.5pt}\/\lor{}\hspace{-14.9pt}\/\vee}}%{\larger${{\mathsf{w}}}$\smaller}%{$\lor\hspace{-2pt}\lor$}

\title{Entanglement and the Path Integral}

\author{Ken Wharton and Raylor Liu \\ 
Department of Physics and Astronomy \\
San Jos\'{e} State University, San Jos\'{e}, CA 95192-0106}
\date{}
\begin{document}
\maketitle

\begin{abstract}
The path integral is not typically utilized for analyzing entanglement experiments, in part because there is no standard toolbox for converting an arbitrary experiment into a form allowing a simple sum-over-history calculation.  After completing the last portion of this toolbox (a technique for implementing multi-particle measurements in an entangled basis), some interesting 4- and 6-particle experiments are analyzed with this alternate technique.  While the joint probabilities of measurement outcomes are always equivalent to conventional quantum mechanics, differences in the calculations motivate a number of foundational insights, concerning nonlocality, retrocausality, and the objectivity of entanglement itself.
\end{abstract}

\section{Introduction}

The observable correlations in entanglement experiments are traditionally calculated using an instantaneous entangled state $\ket{\psi}$.  The mathematical procedure by which joint probabilities are extracted from this state looks explicitly nonlocal in a number of ways.  To the extent that the measured eigenstates $\ket{a}$ and $\ket{b}$ are associated with separated physical systems, the application of the Born rule, $Prob(a,b)=|\bra{\psi}(\ket{a} \otimes \ket{b})|^2$, is a single calculation spanning distant spatial locations.  Furthermore, there is no obvious way to think about $\ket{\psi}$ as a physical intermediary between these locations, because entangled states have no representation in ordinary space (they are vectors in Hilbert space, corresponding to functions on a multi-dimensional configuration space).

An alternative to this traditional calculation is the path integral, which can recover the identical joint probabilities as the state-based analysis \cite{tyagi2022}.  The mathematics of the path integral is less evidently nonlocal.  Instead of using an unlocalized ``state'', the central elements of the path integral are a set of physical histories, and each history always has an evident representation consisting of particle paths connecting distant locations through ordinary spacetime.  In the above example, instead of being applied at the same step in the calculation, the measured result $\ket{a}$ would only be imposed as a boundary constraint on the paths which physically end at that one location, while $\ket{b}$ would only be imposed on different local paths.  Solving the whole problem ``all-at-once'' on the connected zigzag paths through spacetime then yields the correct probabilities.

The aim of this paper is to use the path integral to generate novel perspectives on interesting 4- and 6-particle entanglement geometries which have so far only been analyzed via state-based techniques.  As we shall see, looking at the same experiment through a different mathematical lens can lead to quite different conclusions.  In section 3, a path-integral-based comparison of entanglement swapping \cite{pan1998} and delayed-choice entanglement swapping \cite{peres2000,ma2012} reveals that these two scenarios are not nearly as distinct as a conventional state-based analysis would have one believe.  Then the path integral is applied to the interesting ``triangle network'' \cite{renou2019}, which conventionally appears to be a new sort of nonlocality, but does not seem any more nonlocal than other entanglement scenarios in the path-integral viewpoint.

Both of these applications require making two-particle joint measurements in an entangled basis, a scenario not covered in the recent paper applying the path integral to entangled states \cite{tyagi2022}.  A generic framework for incorporating ``entangled measurements'' into the path integral will therefore first be developed in Section 2, along with a summary of previous techniques for calculating joint probabilities in a path integral framework.  

The particular applications covered in Section 3 are certainly not the only entanglement experiments which could be analyzed from this alternate perspective.  Although the calculated probabilities are just as expected, the ability to take a generic entanglement scenario and analyze it via physical paths in ordinary spacetime seems likely to enhance our understanding of entanglement phenomena.  At minimum, it would help to counterbalance the common belief that entanglement correlations directly link spacelike-separated events, with no physical mediators.  The promise of the path-integral viewpoint is to highlight the uncontroversial connections between such events: the particle worldlines which link everything together in space and time.  It is hoped that such an ``all-at-once'' viewpoint of these connections would provide inspiration for novel approaches to quantum foundations.

\section{The Path Integral Framework}

The essential idea behind the path-integral approach to quantum entanglement is that there are several possible worldline-based ``histories'' which could connect particle sources to the eventual detector outcomes.  Because we are interested in multiparticle situations, we distinguish the term ``path'' from ``history''; a path corresponds to a worldline of a single particle, while a history corresponds to a particular set of paths of all the particles.  Each of these histories can be assigned a complex amplitude $\mathcal{E}$, calculated via some simple rules laid out in the next subsection.  Finally, all of the histories which lead to the same outcomes have their amplitudes first added and then squared to find the (joint) probability of the outcomes in question.  

For simplicity, we will assume that the entangled systems are comprised of photons rather than massive particles.  Also, the entangled photons are assumed to be on well-defined paths, such that one need merely perform a simple sum-over-histories rather than a literal path ``integral''.  

\subsection{Rules for calculating amplitudes}

It is not difficult to calculate the amplitude of a given history, using the basic approach utilized in previous papers \cite{tyagi2022,sinha1991}, especially if one uses ``which-way'' entanglement.  Polarization-based entanglement can be easily converted into which-way entanglement simply by passing the photons through a polarizing beamsplitter, but this will not be required in the examples here.  Consider the basic entanglement geometry in Figure 1, where the vertical axis is representing both a spatial dimension and also time. (This diagram is designed such that photons always travel upward at a 45 degree angle, making these two dimensions coincide.)  The source of two entangled photons, represented as a black diamond, is considered to emit two photons either on one pair of paths (two solid lines) or on another pair of paths (two dashed lines).  Previous work \cite{tyagi2022} indicates how to implement unequal probabilities of these two options, but in the examples below one need only consider equal probabilities, so each pair of paths is ascribed a base amplitude of $1/\sqrt{2}$.  Such a source corresponds to the quantum state $\ket{\psi}$, where
\begin{equation}
    \ket{\psi}=\frac{1}{\sqrt{2}} \ket{solid}_1 \ket{solid}_2 + \frac{1}{\sqrt{2}}\ket{dashed}_1 \ket{dashed}_2.
\end{equation}

There are only two elements which can contribute to the amplitude of a given path: phase plates (circles) and beamsplitters (hollow rectangles -- the solid rectangles are merely mirrors).  The phase plates add a phase delay angle $\beta$ to that particular path, and the corresponding amplitude term is $exp(i\beta)$.  (All other paths shown on these diagrams are assumed to be exactly an integer number of wavelengths, such that no additional phase delays are hidden in the diagrams.)  The beamsplitters allow for two output paths, either a transmission (which picks up an amplitude factor of $\sqrt{T}$), or a reflection (which picks up an amplitude factor of $i\sqrt{R}$).  Here $R$ and $T$ are the (intensity) reflection and transmission coefficients, such that $R+T=1$.

\begin{figure}[ht]
\centering
\includegraphics[width=6cm]{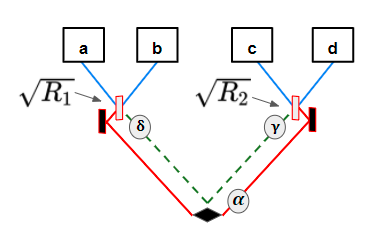}
\caption{Path-integral geometry for a basic entanglement experiment. Detectors a, b, c, and d are placed at the end of potential paths taken by two which-way entangled photons generated from the black diamond source. (If directly measured, the photons are both found on solid line paths or both found on dashed line paths.)  $R_1$ and $R_2$ are the reflectivities of the two beamsplitters. Phase plates, labeled by circles, add the specified phase angle to that path.}
\label{fig:1} 
\end{figure}

Recall that a history consists of a set of paths; the amplitude of the full history is the \textit{product} of the amplitude for each of these paths.  For the two-photon example shown in Figure 1, each history will have two paths, beginning on either both the solid lines or both the dashed lines.  Given the two further possibilities at each beamsplitter, either of these initial cases can end up at any pair of detectors.  For example, the detectors ``$a$'' and ``$d$'' can both fire in one history where the photons are initially on the solid lines and then both reflect from the beamsplitters.  The amplitude of this history is

\begin{equation}
    \mathcal{E}^{ad}_1=\frac{1}{\sqrt{2}} \left( i\sqrt{R_1} \right) \left( e^{i \alpha} i\sqrt{R_2} \right).
\end{equation}
After the base amplitude, the first term is due to the path on the left, and the second term is due to the path on the right (which also passes through the $\alpha$ phase plate).  

But this is not the only possible history that ends at these same two detectors.  Another history begins on the dashed lines, and then transmits through both beamsplitters, yielding
\begin{equation}
    \mathcal{E}^{ad}_2=\frac{1}{\sqrt{2}} \left( e^{i \delta} \sqrt{T_1} \right) \left( e^{i \gamma} \sqrt{T_2} \right).
\end{equation}

Finally, the probability of a joint measurement at detectors $a$ and $d$ is found by adding and squaring the amplitudes,
\begin{equation}
    P(a,d) = \left| \mathcal{E}^{ad}_1 + \mathcal{E}^{ad}_2 \right| ^2.
\end{equation}
Such a probability will always match the corresponding Born rule calculation if the experimental geometry is properly mapped onto the quantum formalism, as has been laid out in detail elsewhere \cite{tyagi2022}.  In this particular case, if each initial solid line is defined to be the state $\ket{0}$ and each initial dashed line is defined to be the state $\ket{1}$, then if one lumps the $\alpha$ phase plate into the preparation procedure, then the example in Figure 1 shows the joint measurement of the state
\begin{equation}
    \ket{\psi}=\frac{1}{\sqrt{2}} e^{i \alpha} \ket{0} \otimes \ket{0} + \frac{1}{\sqrt{2}}\ket{1} \otimes \ket{1}.
\end{equation}
It transpires that the geometry shown in Figure 1, with a measurement by detectors $a$ and $d$, corresponds to the state
\begin{equation}
    \ket{a} \otimes \ket{d}= \left( \sqrt{R_1} \ket{0} + i e^{-i \delta} \sqrt{T_1} \ket{1} \right) \otimes  \left( \sqrt{R_2} \ket{0} + i e^{-i \gamma}\sqrt{T_2} \ket{1} \right),
\end{equation}
from which one can calculate that the path integral approach does indeed give the correct joint probability.

\subsection{Two-Photon Joint Measurements}

For the applications considered below, and many other scenarios, single-qubit measurements on each entangled photon are not sufficient -- we would also like to be able to measure a pair of photons together in an arbitrary entangled basis.  In principle, this can be accomplished in a path-integral-friendly manner using the idea of second harmonic generation (SHG), where two different photons enter a non-linear crystal and produce a single higher-energy photon as output.  (This is literally the time-reverse of the parametric down-conversion process which is commonly used to generate lower-frequency pairs of entangled photons from a single high-frequency pump photon.)

Consider that each qubit in the above example is represented by a pair of paths, one (solid) corresponding to $\ket{0}$ and another (dashed) corresponding to $\ket{1}$.  By crossing the two paths from one qubit with the two paths from another qubit, and putting an SHG crystal at each potential intersection, these two photons can be jointly measured in the computational basis ($\ket{00},\ket{01},\ket{10},\ket{11}$) simply by looking at the second-harmonic emission from each of these crystals with four single-photon detectors, as shown in Figure 2.  A measurement at any one of these four detectors would then (in the path integral view) imply that two photons had in fact crossed paths in the corresponding SHG crystal.

\begin{figure}[ht]
\centering
\includegraphics[width=6cm]{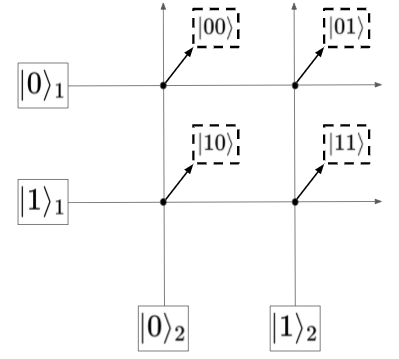}
\caption{The basic two-photon measurement scheme. Black circles represent second-harmonic generating crystals, such that the dashed-box single-photon detectors will only fire if two incoming photons intersect at the corresponding crystal.}
\label{fig:2} 
\end{figure}

Notice that this set-up is not meant to be an actual experiment, but rather an in-principle physical realization amenable to path integral analysis.  In reality, the efficiency of single-photon-input SHG crystals would likely be too low to be useful, but for purposes of the path integral computation we will assume that these devices are somehow 100\% perfect.  The important thing is not the experimental feasibility, but rather a simple path-based account of where the photons are travelling in any given history.

For joint measurements in some other basis, such as the Bell basis, one must remove the four detectors from Figure 2 and send the four paths into an interferometer, as shown in Figure 3.  (Here the dashed boxes are now inputs, representing the second harmonic photon arriving at the corresponding box in Figure 2.)  Using adjustable phase plates and beamsplitters with variable reflectivities, a wide variety of two-qubit bases can be chosen using this geometry.  The only detail that prevents Figure 3 from being completely general is that a few relative phases can only be varied in pairs.  (This could be made fully general by passing the $B$ output into a fifth phase plate and then combining it with the $C$ output in a fifth beamsplitter, but we are not aware of any useful basis for which these additional elements would be required.)

\begin{figure}[ht]
\centering
\includegraphics[width=8cm]{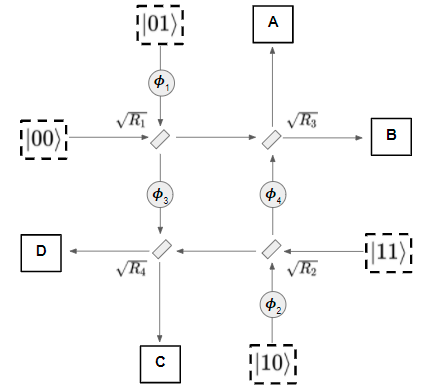}
\caption{Implementation of a two-photon entangled measurement device; the initial input stage from Figure 2 is not shown.   Four phase plates and four adjustable-reflectivity beamsplitters can potentially affect the second-harmonic photon before arriving at the $A$, $B$, $C$, and $D$ single-photon detectors.}
\label{fig:3} 
\end{figure}

The precise measurement basis can be found from Figure 3 by propagating the four input kets to the four detectors.  The amplitudes and phases of each of these terms are affected using rules that entirely mimic the path integral procedures defined above. (This is only because we are down to a single photon at this point; recall that after the SHG crystals, even with two original photons there is only one second harmonic output photon, on one of four possible paths.) Evolving these inputs through the interferometer then yields the general relationships

%\begin{eqnarray}
%A&=&ia\sqrt{T_1R_3}-be^{i\phi_1}\sqrt{R_1R_3}+ce^{i\phi_4+i\phi_2}\sqrt{T_2T_3}+de^{i\phi_4}i\sqrt{R_2T_3}\\
%B&=&\sqrt{T_1T_3}a+ie^{i\phi_1}\sqrt{R_1T_3}b+ie^{i\phi_2+i\phi_4}\sqrt{T_2R_3}c-e^{i\phi_4}\sqrt{R_2R_3}d\\
%C&=&ie^{i\phi_3}\sqrt{R_1T_4}a+e^{i\phi_1+i\phi_3}\sqrt{T_1T_4}b-e^{i\phi_2}\sqrt{R_2R_4}c+i\sqrt{T_2R_4}d\\
%D&=&-e^{i\phi_3}\sqrt{R_1R_4}a+ie^{i\phi_1+i\phi_3}\sqrt{T_1R_4}b+ie^{i\phi_2}\sqrt{R_2T_4}c+\sqrt{T_2T_4}d
%\end{eqnarray}

\begin{align}
&\begin{aligned}
\ket{A}&= ie\sqrt{T_1R_3}\ket{00}-e^{i\phi_1}\sqrt{R_1R_3}\ket{01}\\
&\qquad +e^{i\phi_2+i\phi_4}\sqrt{T_2T_3}\ket{10}+ie^{i\phi_4}\sqrt{R_2T_3}\ket{11}\end{aligned} \\
&\begin{aligned}
\ket{B}&= \sqrt{T_1T_3}\ket{00}+ie^{i\phi_1}\sqrt{R_1T_3}\ket{01}\\
&\qquad +ie^{i\phi_2+i\phi_4}\sqrt{T_2R_3}\ket{10}-e^{i\phi_4}\sqrt{R_2R_3}\ket{11}\end{aligned}\\
&\begin{aligned}
\ket{C}&= ie^{i\phi_3}\sqrt{R_1T_4}\ket{00}+e^{i\phi_1+i\phi_3}\sqrt{T_1T_4}\ket{01}\\
&\qquad -e^{i\phi_2}\sqrt{R_2R_4}\ket{10}+i\sqrt{T_2R_4}\ket{11}\end{aligned}\\
&\begin{aligned}
\ket{D}&= -e^{i\phi_3}\sqrt{R_1R_4}\ket{00}+ie^{i\phi_1+i\phi_3}\sqrt{T_1R_4}\ket{01}\\
&\qquad +ie^{i\phi_2}\sqrt{R_2T_4}\ket{10}+\sqrt{T_2T_4}\ket{11}\end{aligned}
\end{align}

Therefore, by choosing the appropriate beamsplitters and phase plates, almost any desired measurement basis can be implemented, and it will effectively be an ``entangled measurement'' of the original two-qubit state.  

%For example, if one wanted to specifically measure the original two photons in the Bell basis (as will be done in the next section), one would set the reflectivities $R_1=1$, $R_2=R_3=\frac{1}{2}$, and $R_4=0$. Additionally, the phase plates would all be set to zero, as the beamsplitter arrangement with the factors of $i$ picked up from reflections naturally yield the desired Bell basis. 

%We can further simplify from the PBR basis into the Bell basis by setting $\phi_3=0$ and $R_4=0$.

\section{Examples}

\subsection{Entanglement Swapping}

By extending the path integral framework to include two-photon entangled measurements, a wide array of intriguing entanglement phenomena becomes amenable to a path-based analysis.  One of the more surprising consequences of quantum theory (apart from entanglement itself) is the case of ``entanglement swapping'' (ES), where two photons are said to become entangled despite never interacting with each other in the past or future \cite{pan1998}.  Indeed, entanglement can even be arranged to occur between two photons whose worldlines are entirely spacelike-separated from each other.

The essential geometry behind ES is shown in Figure 4.  Here two different pairs of entangled photons (each created by its own two-photon source) are arranged in a $\W$ shape in spacetime.  After a joint measurement of the two photons at the central vertex (in the Bell basis), conventional QM requires updating the state of the two photons on the outer wings of the $\W$ to be in an entangled configuration.  The question of \textit{which} entangled state depends on the outcome of the central measurement; four different entangled states are therefore possible.  Such an experimental geometry has been carried out in a number of contexts, and indeed the photons on the wings obey the expected Bell correlations, given the known outcome at the central vertex.  

\begin{figure}[ht]
\centering
\includegraphics[width=8cm]{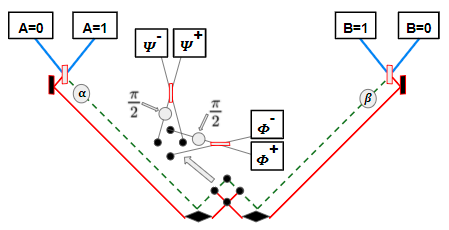}
\caption{An entanglement swapping (ES) geometry useful for path-integral analysis. Each black diamond is a source of two which-way entangled photons.  The central measurement device uses four SHG crystals (black circles), drawn twice for clarity.  All four beamsplitters (hollow rectangles) are $50/50$; two phase plates add a fixed $\pi/2$ phase, the other two ($\alpha,\beta$) are adjustable.  Four central detectors (for single second-harmonic photons) correspond to the four Bell states. Alice's two single-photon detectors are on the left wing; Bob's are on the right wing; these are sensitive to the original first-harmonic photons.}
\label{fig:4a} 
\end{figure}

It is important to emphasize that the entanglement on the wings is only evident if the central outcome is known.  If averaged over \textit{all} central outcomes, the four different entanglement correlations happen to cancel out.  In that case no entanglement can be inferred from the data alone -- indeed, no correlations of any sort survive such a marginalization over the central measurement.  (If this were not true, it would be possible to nonlocally signal from the central measurement to the wings, if the wings were brought together.)

In Figure 4, the ES scenario is diagrammed in a path-integral-friendly geometry.  As before, each of the two photon sources emit photons in matching modes -- either on both solid lines ($\ket{00}$), or on both dashed lines ($\ket{11}$).  (The sources are independent; the two modes on the left do not have to match the two modes on the right.)  This corresponds to the four-photon preparation 
\begin{equation}
    \ket{\psi_{4}} = \frac{1}{2} \ket{0000} + \frac{1}{2} \ket{0011} + \frac{1}{2} \ket{1100} + \frac{1}{2} \ket{1111}.
\end{equation}
The notation here reads left-to-right on Figure 4, such that $\ket{0011}$ corresponds to the left source emitting on the solid lines and the right source emitted on the dashed lines.

The two center photons converge on one of the four SHG crystals.  To keep the diagram legible, these crystals are drawn twice.  The single output photon (second harmonic, represented by a thin solid line) either reflects or transmits through a 50/50 beamsplitter before being detected at one of four single-photon detectors.  This geometry is chosen to match the standard Bell basis.  It is simple to check that the device pictured would be sensitive to the states
\begin{eqnarray}
\ket{\Psi^+}&=&\frac{1}{\sqrt{2}}\left( \ket{01} + \ket{10} \right)  \\
\ket{\Psi^-}&=&\frac{1}{\sqrt{2}}\left( \ket{01} - \ket{01} \right) \nonumber \\
\ket{\Phi^+}&=&\frac{1}{\sqrt{2}}\left( \ket{00} + \ket{11} \right) \nonumber \\
\ket{\Phi^-}&=&\frac{1}{\sqrt{2}}\left( \ket{00} - \ket{11} \right).\nonumber
\end{eqnarray}
This notation refers to the middle two photons reading left to right, so the full state $\ket{0011}$ would correspond to $\ket{01}$ for this measurement.

After this measurement in the center, Alice (on the left) and Bob (on the right) measure the single photon on their wing of the experiment.  This is done in the identical manner as in Figure 1, although their beamsplitters are fixed to 50/50 for simplicity.  They can still vary their local setting through their corresponding phase plate; Alice can set the phase angle $\alpha$ and Bob can set the phase angle $\beta$.  Alice's outcome is $A$, and Bob's outcome is $B$.  

Suppose the measurement in the center corresponds to the $\ket{\Phi^+}$ outcome.  Using conventional QM, this collapses $\ket{\psi_{4}}$, leaving the state of Alice and Bob's photons in the maximally-entangled combination
\begin{equation}
\label{eq:psiint}
    \ket{\psi_{int}} = \frac{1}{\sqrt{2}} \left( \ket{0}_A \otimes \ket{0}_B + \ket{1}_A \otimes \ket{1}_B \right).
\end{equation}
In conventional QM, one would then take this intermediate state $\ket{\psi_{int}}$, and apply the Born rule with Alice and Bob's measured bases to determine the outcome probabilities.  For example, in Figure 4, the joint outcome $A=0$ and $B=0$ would correspond to
\begin{equation}
\label{eq:a0b0}
    \ket{A=0,B=0} = \frac{1}{\sqrt{2}} \left( \ket{0}_A + ie^{-i\alpha}\ket{1}_A \right) \otimes \frac{1}{\sqrt{2}} \left( \ket{0}_B + ie^{-i\beta}\ket{1}_B \right).
\end{equation}
The Born rule then yields the usual Bell correlations.

But in the path-integral approach, this two-step process (passing through $\ket{\psi_{int}}$) is not an option.  The only way to calculate the probabilities is ``all at once'', considering the central measurement outcome $C$ on the same footing as Alice and Bob's measurement.  To calculate the joint probabilities $P(C=\Phi^+,A,B)$, note that each joint outcome can only have two possible histories. For $\ket{\Phi^+}$, we can trace backwards to see that all photons must be on the solid lines or all must be on the dashed lines.

As before, we can calculate the amplitude of each of these two histories, for a given set of outcomes. For example, the all-dashed-line history for $P(\Phi^+,A=0,B=0)$ gives an amplitude of 
\begin{equation}
    \mathcal{E}^{\Phi^+\!\!,0,0}_1=\tfrac{1}{2}\left(e^{i\alpha}\tfrac{1}{\sqrt{2}}\right)\left(\tfrac{i}{\sqrt{2}}\right)\left(e^{i\beta}\tfrac{1}{\sqrt{2}}\right),
\end{equation}
 where the $1/2$ is the base amplitude of the two sources.  Using the above rules, the first term comes from Alice's photon, the middle term comes from the central second harmonic photon (passing through the $\pi/2$ phase plate and then the 50-50 beamsplitter), and the final term is Bob's photon.  
 
 Meanwhile, the all-solid-line history for $P(\Phi^+,A=0,B=0)$ is
\begin{equation}
    \mathcal{E}^{\Phi^+\!\!,0,0}_2=\tfrac{1}{2}\left(\tfrac{i}{\sqrt{2}}\right)\left(\tfrac{i}{\sqrt{2}}\right)\left(\tfrac{i}{\sqrt{2}}\right).
\end{equation}
We can then sum these two amplitudes and take the magnitude squared to find the joint probability.  This result, along with the other three joint outcomes associated with $P(\Phi^+)$, can be calculated as
\begin{align}
&\begin{aligned}
\label{eq:first}
P(\Phi^+,A=0,B=0)=\tfrac{1}{32}\left|e^{i(\alpha+\beta)}-1\right|^2,\end{aligned}\\
&\begin{aligned}
P(\Phi^+,A=0,B=1)=\tfrac{1}{32}\left|e^{i(\alpha+\beta)}+1\right|^2,\end{aligned} \nonumber \\
&\begin{aligned}
P(\Phi^+,A=1,B=0)=\tfrac{1}{32}\left|e^{i(\alpha+\beta)}+1\right|^2,\end{aligned} \nonumber \\
&\begin{aligned}
P(\Phi^+,A=1,B=1)=\tfrac{1}{32}\left|e^{i(\alpha+\beta)}-1\right|^2.\end{aligned}\nonumber
\end{align}

For this geometry, there are 12 other possible joint outcomes -- four possible Alice/Bob results each corresponding to the other three possible central measurements.  (They are not shown here, but are easily calculated using the path integral approach.)  All 16 of these joint probabilities add to one.  However, if one knew that the result of the central measurement corresponded to $\ket{\Phi^+}$, one would of course use this additional knowledge to Bayesian-update these other 12 probabilities to zero.  Renormalizing the above four expressions (multiplying these probabilities by 4) thereby yields exactly the same probabilities one would calculate from the Born rule applied to $\ket{\psi_{int}}$ (\ref{eq:psiint}).  For instance, the inner product of $\ket{\psi_{int}}$ with $\ket{A=0,B=0}$ (\ref{eq:a0b0}) yields the term proportional to the first expression in (\ref{eq:first}).  Even though there is no intermediate step in the path integral calculation, the predictions are still precisely the same as in conventional QM.

From this result, some readers may draw the conclusion that the path integral account cannot tell us anything interesting that could not have been already deduced from conventional QM.  After all, the joint probabilities are identical, and those are the only things measurable in the laboratory; all other considerations might be thought as irrelevant.  Nevertheless, the next example will demonstrate that the the formalism one chooses can have intriguing consequences.

\subsection{Delayed Choice Entanglement Swapping}

The relative timing of the measurements made in an entanglement swapping experiment can be altered in a seemingly simple way, transforming the original Figure 4 geometry into the ``delayed choice entanglement swapping'' (DCES) configuration shown in Figure 5 \cite{peres2000,ma2012}.  Again, two pairs of entangled photons are created, and again Alice measures the left-most photon and Bob measures the right-most photon.  Also, as before, the central two photons are measured in the standard Bell basis, using the same configuration as in Figure 4.  The only difference is one of relative timing; in DCES, Alice and Bob make their measurements first, and the central Bell-basis measurement is made last.

The path-integral account of this experiment is identical to the analysis of ordinary entanglement swapping in every respect.  Again, there are 16 possible joint outcomes, and the probability of every single one of these 16 outcomes follows precisely the same calculation as before.  Even though the histories in question have a slightly different $\W$ shape in spacetime, the relative timing of the measurements does not affect either the amplitudes or the probabilities.  And indeed, experimental results bear this out: the joint probabilities in DCES (Figure 5) are identical to the joint probabilities in ordinary entanglement swapping (Figure 4).

\begin{figure}[ht]
\centering
\includegraphics[width=8cm]{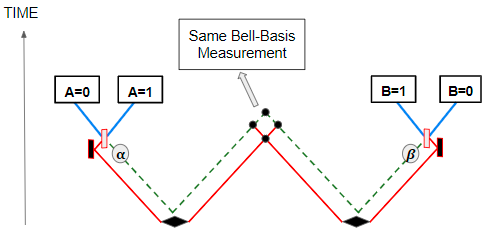}
\caption{The delayed choice entanglement swapping geometry (DCES).  Essentially the same as Figure 4, but with the central Bell-basis measurement made after the measurements in the wings.}
\label{fig:4b} 
\end{figure}

And yet the conventional QM account of DCES does not consider these two experiments to be essentially identical.  In fact, they are widely thought to be quite different \cite{egg2013,price2021}.  While QM takes it to be incontrovertible that Alice and Bob's photons are entangled in Figure 4, it would not conclude that Alice and Bob's photons were entangled in Figure 5.  The appearance of entanglement-like correlations between Alice and Bob's measurement is widely accepted to not be an indication of true entanglement, but rather a post-selection artifact, apparent only after the final measurement is made and then certain runs of the experiment are chosen for analysis.\cite{price2021}  (Recall, if all the data is analyzed, in either version of the experiment, there are no correlations between $A$ and $B$.)

We will not go through all the details of how a conventional QM analysis would apply for Figure 5, but the essential idea is that Alice's measurement result $A$ would collapse the corresponding entangled photon in the center.  A similar collapse would occur for Bob's result, $B$.  The state of the two unmeasured central photons therefore is correlated with the (random, uncorrelated) outcomes $A$ and $B$.  Then, after those central photons are measured in the Bell basis, the result of the central measurement can plausibly be correlated with the particular values of $A$ and $B$.  Post-selecting on one of these central measurements then gives the illusion of entanglement-generated correlations between Alice and Bob, despite the fact that those two photons were never entangled.\cite{egg2013}

The conflict with the path integral perspective will be further discussed in the final section, but already it should be clear that a ``peaceful coexistance'' of these two perspectives is mostly untenable.  Either Figure 4 and Figure 5 are essentially the same experiment, or else they are not.  First, however, the path integral will be applied to an intriguing case concerning three pairs of entangled particles.

\subsection{The Triangle Network}

As a final application, a recent result \cite{renou2019} has indicated that a ``new form of quantum nonlocality'' can emerge in an experimental geometry called the triangle network.  The correlations in question are distinct from Bell-type inequalities because they do not require the variation of measurement settings.  In other words, the settings in this network can stay fixed, and yet still produce a probability distribution which is classically impossible (assuming the usual restrictions on hidden variable models \cite{wharton2020}).  

Consider the geometry shown in Figure 6.  Unlike the ES and DCES diagrams, this geometry requires two distinct spatial dimensions, so time is no longer mapped to the vertical axis. (Meaning, photons do not need to propagate upward.)  Three sources of entangled photon pairs are represented by the small black diamonds, one in the center of each side of the triangle.  Each source emits two photons, either on the two solid lines or the two dashed lines.  As before, each of these sources corresponds to the state $\sqrt{2}^{-1}(\ket{00}+\ket{11})$.  (We will continue to assign $\ket{0}$ to solid lines and $\ket{1}$ to dashed lines.)  At each of three identical measurement stations (A,B, or C), two photons from two different sources are jointly measured in the basis shown, using the same SHG crystals as before.  The beamsplitters are no longer 50-50, but each one has a transmission factor of $\sqrt{T}=u$, and a reflection factor of $\sqrt{R}=v$, such that $u^2+v^2=1$.

From the measurement geometry, one can ascertain that the $X$ outcome for each measurement corresponds to the incoming state $\ket{01}$, and the $Y$ outcome corresponds to the incoming state $\ket{10}$.  (The convention here is that looking back at the photon sources, each experimenter labels the photons in left-to-right order, such that $\ket{01}$ implies a solid line input coming from the left, and a dashed line input coming in from the right.)  Because the $\pi/2$ phase plate introduces the same phase as a reflection off the beamsplitter, the other two possible joint measurement outcomes $W$ and $Z$ correspond to the states
\begin{eqnarray}
 \ket{W} &=& v \ket{00} - u \ket{11} \\
 \ket{Z} &=&  u\ket{00}+v\ket{11}.
\end{eqnarray}

\begin{figure}[ht]
\centering
\includegraphics[width=9cm]{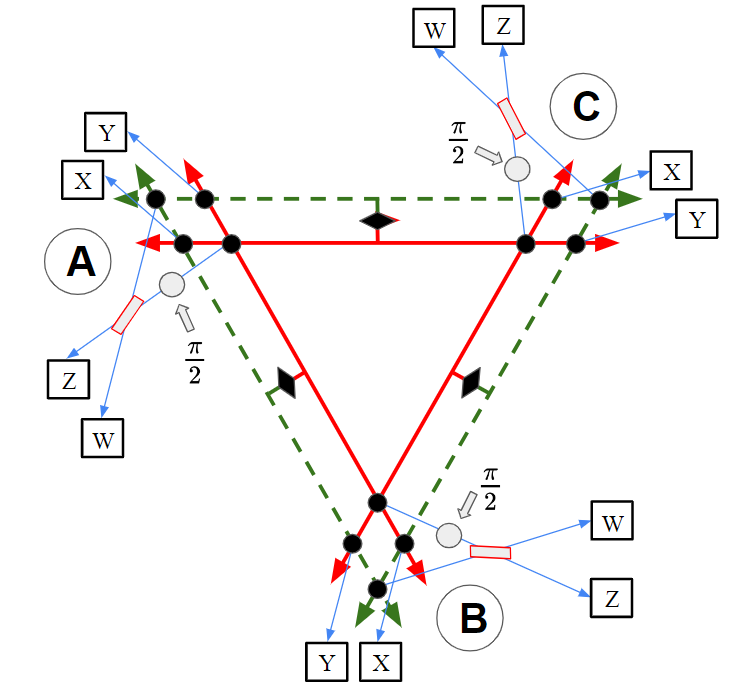}
\caption{The triangle network geometry.  Unlike Fig. 4 and Fig. 5, time is not mapped to the vertical axis; the spacetime shape of this experiment looks like $\VW$.  Each black diamond is a which-way-entangled two photon source.  Three measurement stations ($A,B,C$) each use four SHG crystals (black circles) and four single-photon detectors ($W$, $X$, $Y$, or $Z$) sensitive to second-harmonic photons.  Each station also uses a $\tfrac{\pi}{2}$ phase plate and an unlabeled $50/50$ beamsplitter.}
\label{fig:6} 
\end{figure}

For one example, consider the probability $P(W_A,W_B,Z_C)$, meaning that measurement stations $A$ and $B$ make a measurement of $W$ while station $C$ makes a measurement of $Z$. We can first rule out the possibility of having a history with photons on both solid and dashed lines. If we were to have a combination of solid and dashed lines, we would necessarily find ourselves with some measurements at $X$ or $Y$, which is in conflict with the $(W,W,Z)$ outcome we are calculating. Therefore we are only interested in histories for which the sources emit either along all dashed lines or all solid lines. This will lead to two different amplitudes corresponding to the $(W,W,Z)$ outcome. 

For the amplitude of the all-solid-line history, we see SHG photons at stations $A$ and $B$ pass through a $\pi/2$ phase plate and undergo a reflection before arriving at detector $W$. Each of these photons will pick up an amplitude of $i^2v$, combining the factor $iv$ from reflection and the factor $i$ from the phase plate. At station $C$, we see a transmission ($u$) of a SHG photon to detector $Z$ with the same $\pi/2$ phase plate. Therefore, the photon at $C$ will pick up a factor $iu$. Recall that each of these three photon sources comes with an amplitude of $1/\sqrt{2}$, so altogether

\begin{equation}
    \mathcal{E}^{W,W,Z}_{solid}=\frac{1}{2\sqrt{2}}(i^2v)(i^2v)\left(iu\right)=\frac{iuv^2}{2\sqrt{2}}.
\end{equation}
Likewise, applying the same analysis for the dashed-line history, one has two beamsplitter transmissions and one reflection:
\begin{equation}
    \mathcal{E}^{W,W,Z}_{dashed}=\frac{1}{2\sqrt{2}}\left(u\right)\left(u\right)\left(iv\right)=\frac{iu^2v}{2\sqrt{2}}.
\end{equation}
To find the probability of this outcome we sum the amplitudes of these two histories and take the magnitude squared:
\begin{equation}
    P(W_A,W_B,Z_C)=\frac{1}{8}\Big|\left(iuv^2\right)+\left(iu^2v\right)\Big|^2=\frac{1}{8}u^2v^2(u+v)^2.
\end{equation}
Comparing this result to the original QM calculation \cite{renou2019}, this in indeed the joint probability of measurements made by detectors $W$, $W$, and $Z$ at respective stations $A$, $B$, and $C$. 

Like the worked example above, other possible outcomes also have two histories, each with all-solid or all-dashed line photon emissions.  These cases are:
\begin{align}
&\begin{aligned}
P(W_A,W_B,W_C) &=& \tfrac{1}{8}\Big|(i^2v)(i^2v)(i^2v)+\left(u\right)\left(u\right)\left(u\right)\Big|^2\end{aligned}
\\
&\begin{aligned}
P(Z_A,Z_B,Z_C) &=& \tfrac{1}{8}\Big|\left(iu\right)\left(iu\right)\left(iu\right)+\left(iv\right)\left(iv\right)\left(iv\right)\Big|^2\end{aligned}
\nonumber \\
%&\begin{aligned}
%P(W_A,W_B,Z_C) &=& \tfrac{1}{8}\Big|(i^2v)(i^2v)\left(iu\right)+\left(u\right)\left(u\right)\left(iv\right)\Big|^2\end{aligned}
%\nonumber \\
&\begin{aligned}
P(W_A,Z_B,Z_C) &=& \tfrac{1}{8}\Big|(i^2v)\left(iu\right)\left(iu\right)+\left(u\right)\left(iv\right)\left(iv\right)\Big|^2\end{aligned}
\nonumber
\end{align}

 Note that because of the rotational symmetry in the triangle network, $120$-degree rotations produce identical geometry and accordingly, identical probabilities.  Given this, the above cases cover the all-dashed-histories and all-solid-histories, but there can be other combinations which allow some $X$ or $Y$ outcomes.  These cases will only have one history (or zero), never two.  For example, consider $P(W_A,X_B,Y_C)$. In order to make such a measurement, only one history is possible. Specifically, one of the sources must emit along the dashed-line paths (towards $B$ and $C$), and the other two sources must emit along the solid line paths. The amplitude calculation for this single history is simply
\begin{equation}
    \mathcal{E}^{W,X,Y}=\frac{1}{2\sqrt{2}}(i^2v)(1)(1).
\end{equation}
Here we have picked up a factor of $iv$ and $i$ through the phase plate and reflection at location $A$ while intersections at $X_B$ and $Y_C$ maintain an amplitude of $1$. Therefore,
\begin{equation}
    P(W_A,X_B,Y_C)=\frac{1}{8}|i^2v|^2=\frac{1}{8}v^2.
\end{equation}

For the remaining joint probabilities with a single possible history:
\begin{align}
%&\begin{aligned}
%P(W_A,X_B,Y_C) &=& \tfrac{1}{8}\Big|i^2v\Big|^2\end{aligned}
%\\
&\begin{aligned}
P(W_A,Y_B,X_C) &=& \tfrac{1}{8}|u|^2\end{aligned} \\
&\begin{aligned}
P(Z_A,X_B,Y_C) &=& \tfrac{1}{8}|iu|^2\end{aligned}
\nonumber \\
&\begin{aligned}
P(Z_A,Y_B,X_C) &=& \tfrac{1}{8}|iv|^2\end{aligned}.
\nonumber
\end{align}

Apart from permutations of the above cases, all other measurement combinations are prohibited due to the geometry of the setup. For instance, any joint measurements which do not go through beamsplitters (that is, only detectors $X$ and/or $Y$ make a measurement) require the intersection of a solid line and a dashed line. This simply can't be the case as at least two of the sources must be the same type of line. Similarly, other cases like $P(W_A,X_B,X_C)$ are also forbidden since the detectors perceive the sources on a left-to-right convention. $X_C$ perceiving a solid line from the left implies $X_B$ is not possible. (Such a result would imply that the source between $B$ and $C$ would have emitted photons on one solid and one dashed line.)  And any joint measurement that has at least two $W$ or $z$ measurements necessarily requires another $W$ or $z$ measurement for the remaining detector, as they imply an all-dashed-line or all-solid-line history. This means that measurements like $P(W_A,Z_B,Y_C)$ are impossible.

It is worthwhile to note that this reasoning for the impossible joint outcomes is entirely based classical logic, given the experimental geometry.  This is in general agreement with the analysis of other impossible joint outcomes in an upcoming paper\cite{WhartonAdlam}, but this logic may not be strictly evident from a conventional QM calculation.  Nevertheless, all of the probabilities calculated here, using the path integral approach, are identical to the probabilities for this scenario as calculated by conventional QM \cite{renou2019}.

\section{Analysis}

Many physicists view path integrals as merely an alternate technique, provably equivalent to conventional QM in every way.  From such a perspective, it should not matter which method one uses to calculate the joint probabilities of an entanglement experiment; the two approaches would always yield the same result.  And indeed, the above results bear out this operational equivalence.

But this apparently ``peaceful coexistence'' seems to go no further than the bare probabilities.  To the extent that entanglement experiments are interesting, it is presumably because the experimentally-measured probabilities allow us to infer certain features of reality, features that go beyond the operational predictions of a black-box model.  And these inferences appear to be substantially different, depending on how the probabilities are calculated.

The analysis in this section will detail clear differences between the implications of these two techniques.  They will lead to different conclusions about a wide variety of topics, ranging from the appearance of nonlocality to the question of whether two systems can be ``entangled'' in a fully objective sense.

\subsection{States vs Histories}

One essential difference between path integrals and conventional QM, highlighted by the above examples, is that the former calculates the probabilities of entire measurement histories, while the latter calculates the probability of temporally-localized measurement events.  For a chain of such events, conventional QM necessarily uses a chain of calculations, alternating the Born Rule with a state-update (collapse) rule, repeating as needed to generate the conditional probabilities.  (These can then be combined to generate the joint probabilities, if desired.)  

In conventional QM, when two space-like-separated measurements are made on an entangled system, it is possible to treat these measurements as if they are made simultaneously and calculate the joint probability in one step.  But it is generally understood that this should not be done for time-like separated measurements.  This issue was highlighted in the Section 3.2 discussion of Delayed Choice Entanglement Swapping (DCES), where the geometry in Figure 5 does not specify whether the central joint measurement is space-like or time-like separated from Alice and Bob.  If the central measurement is in the absolute future of Alice and Bob, there is no possible reference frame where all the measurements are simultaneous, and a sequence of calculations is required by conventional QM.

On the other hand, sequential-calculations are not even \textit{possible} for the path integral.  The mathematics of this approach necessarily solves for the complete joint probability, all at once.  In other words, the path integral assigns probabilities to entire patterns over space and time, not to instantaneous events.  In order to perform the path integral calculations of just the Alice-Bob correlations in Section 3.1, the full joint probabilities first had to be calculated in Eqn. (\ref{eq:first}) before one could Bayesian-update on the known central outcome.  It is notable that when using the path integral the conditional probabilities can only be calculated from the joint probabilities, while in conventional QM it is usually the other way around.

This distinction is certainly important to discussions of whether there is some new sort of ``objective probability'' in quantum theory \cite{adlam2021}, as distinguished from ordinary subjective classical probability (the latter results from not having complete information).  Before that topic could be properly investigated, one would have to clarify which is fundamental: conditional probabilities of states or joint probabilities of entire sequences.  The two techniques discussed here disagree about which one should be derived from the other.

\subsection{Objective Entanglement}

Any peaceful coexistence is further threatened by the differing perspective of entanglement as an objective feature of quantum systems.  Consider the comparison between ordinary entanglement swapping (ES, Figure 4) and delayed choice entanglement swapping (DCES, Figure 5).  The path integral analysis takes these to be essentially the same experiment, with not a single difference in how the probabilities are calculated between these two cases.  But conventional QM takes ES and DCES to be quite different.  ES seems to require that Alice's and Bob's photons are truly entangled, while DCES seems not to have any true entanglement between Alice and Bob's photon (instead explaining those correlations as a product of post-selection \cite{price2021}).  If ``entanglement'' is thought to be an objective property of a system, then these two experiments are inherently different, and the conventional QM perspective cannot be reconciled with the path integral perspective.

The bare correlations cannot distinguish between these views; they are the same in both cases.  Indeed, this might be evidence that somehow the path integral is giving a clearer explanation of the relationship between these two experiments, as the identical joint probabilities would naturally follow if the two experiments were essentially equivalent.

Another argument against the conventional QM perspective is that somehow the existence of ``entanglement'' between Alice's and Bob's photons seems to depend on one's reference frame, as this would determine the precise time-ordering of the three measurements in question.\cite{price2021}  Nevertheless, even with these arguments, we take it to still be conventional wisdom that there really is entanglement between the wings of the experiment in Figure 4, and there really is not entanglement between the wings of the experiment in Figure 5.\cite{egg2013}  If this viewpoint is correct, it must mean the path integral account is essentially mistaken, or at least missing a key piece of physics hiding in instantaneous states.  Either way, the two viewpoints are in stark contrast.

\subsection{Analysis of the Triangle Network}

This brings us to the final example, the triangle network from Section 3.3.  Unlike the previous examples, this case does not have a sequence of measurements, but rather takes all three measurements to occur at once.  A number of open questions remain concerning the precise distinction between triangle-network correlations and traditional Bell-correlations \cite{kraft2021,tavakoli2021}, but it is clear that the triangle-network allows for a new sort of no-go theorem without requiring different settings for different runs of the experiment.  

In a Bell-style no-go theorem, one runs the experiment with one pair of Alice-Bob settings ($\alpha, \beta$), and then runs the experiment again with another pair of settings.  (This could be done for the ES example in Section 3.1, so long as one consistently enforced the same central measurement.)  To derive Bell's no-go result, along with the assumption of no direct space-like influences, the other crucial assumption is Bell Statistical Independence.  This assumption states that the correlation-mediating variables $\lambda$ are independent of the eventual settings, $P(\lambda|\alpha,\beta)=P(\lambda)$.  A violation of the inequalities which follow from these assumptions proves that any model of these phenomena must either allow nonlocality (direct spacelike influences) or future-input dependence (an influence of the settings on the initial values of $\lambda$). \footnote{A third option, superdeterminism, posits a common cause of both $\lambda$ and the settings.}

In contrast to these Bell-type scenarios, the triangle network has a quite different no-go theorem, with no need for adjustable settings.  Along with the assumption of no direct influences between disconnected worldlines, the no-go theorem for the triangle network also assumes independent source statistics for the three pairs of entangled particles.  Specifically, if $\lambda_j$ fully describes the prepared particles at the $j$th source, it seems reasonable to assume that the only reasonable source correlations must take the separable form 
\begin{equation}
    \label{eq:sep}
P(\lambda_1,\lambda_2,\lambda_3)=P(\lambda_1)P(\lambda_2)P(\lambda_3).
\end{equation}
The violation of the resulting ``triangle-locality'' condition (for certain fixed values of $u$; see \cite{renou2019,pozas2022} for details) therefore proves either nonlocality (direct influences between disconnected worldlines) or else some explanation of hidden correlations which violate Eqn. (\ref{eq:sep}).  

It is this latter option that depends on the framework used to calculate the probabilities.  In the conventional state-based QM analysis, the initial state of the 6-photon system is in three separate locations, with no shared past history.  From this viewpoint, it is quite difficult to imagine any violation of Eqn. (\ref{eq:sep}) that was not effectively a nonlocal influence.  For example, if nature made some random choice for the values of $\lambda_1$, and this choice was somehow nonlocally available over at another pair of generated photons, perhaps $\lambda_2$ could then be correlated with $\lambda_1$ in a way that violated Eqn. (\ref{eq:sep}).  But this account would simply be another type of direct space-like influence.  Even a future-input-dependent model would seem to perhaps fail (as discussed further below) because there is no obvious relevance of the future setting in Eqn. (\ref{eq:sep}).

On the other hand, the all-at-once analysis of the path integral naturally looks at the entire history of the experiment.  In spacetime, the six photons are connected in a $\VW$ pattern, where the three final vertices are linked via the experiment in essentially the same manner as the three \textit{initial} vertices.  Viewed holistically, it would be far less difficult to explain a violation of Eqn. (\ref{eq:sep}).  After all, one would not assume that the final vertices would remain uncorrelated from such an interconnection, and by the same argument some global all-at-once rule could correlate certain hidden parameters on the initial vertices.  Given the connections along the photon worldlines, such an explanation would not appear to require direct space-like influence.  In other words, the path integral account of these experiments does not generally treat them as ``nonlocal'', but rather as evidence that the entire histories must be analyzed via all-at-once rules as advocated in \cite{wharton2014,adlam2022b,chen2021}.

Apart from disagreeing about how ``nonlocal'' such experiments might be, the path integral viewpoint may also offer a topological account of why triangle-network correlations seem to be so different from traditional Bell-correlations.  Consider that the entanglement-swapping experiments in Sections 3.1 and 3.2 also have three measurement locations, counting the central measurement.  And yet the correlations in ES and DCES look to be dramatically different from the three-measurement correlations possible in the triangle network topology.\cite{renou2019,tavakoli2021}  Any analysis of this difference based on instantaneous states seems inferior to analyzing the entire histories of the path integral, using the important context of the connectivity pattern in spacetime.  

The path integral uses spacetime topologies such as $\W$ to calculate entanglement swapping, but adds more connections with the $\VW$ topology of the triangle network.  Even classically, adding such additional connections would lead to different correlations.  So instead of categorizing these experiments by their Hilbert space dimensionality, or their bare correlations, a categorization based on the pattern of spacetime connectivity would seem to be much more fruitful.  Further research efforts in this direction should note that there are certain natural symmetries of path integral accounts \cite{WMP}, in agreement with the apparent identity between the $\W$ geometries of Section 3.1 and 3.2.  And since it is evidently impossible to permute a $\W$ topology into a $\VW$ topology, one would naturally expect novel correlation patterns in the latter case.  This would not imply that the latter case must be more ``non-local'', but simply more connected.

\subsection{The Path Integral and Retrocausation}

Because the path integral analyzes entire histories, locally connected by possible particle trajectories, it does not suggest direct spacelike nonlocality.  As outlined earlier in this section (and in \cite{wharton2020}), an alternative to non-local accounts are those which are future-input dependent or ``retrocausal''.

There are reasonable arguments that the path integral is naturally aligned with retrocausal models.  Without any discontinuous collapse, the path integral is much more evidently time-symmetric than conventional QM, bringing it closer to a causally-neutral account.  Furthermore, all path-integral calculations require fixing the final condition along with an initial condition, and final boundary conditions are known to induce the appearance of retrocausal behavior, such as systems with an opposite entropic arrow of time \cite{schulman,wharton2010}.  

However, pushing against these arguments, it remains true that there is no evident interpretation of exactly what the path integral mathematics is implying about the underlying system.  Some efforts to realistically interpret this mathematics indeed look retrocausal \cite{wharton2016}.  But alternatively one might stress that different histories are being combined into the same probability calculation, and then conclude that the path integral was supporting a many-worlds-type picture, with a different sort of ``nonlocal'' interaction between histories/worlds. 

But as a result of the above analysis -- particularly the case of the triangle network -- it is now possible to at least distinguish between two very different categories of retrocausal models, and judge one of them more compatible with the path integral than the other.  Adlam has named these categories ``dynamical retrocausation'' and ``all-at-once retrocausation'' \cite{adlam2022}.

Dynamical retrocausation is a category of models which generally result from equating causation with some dynamical process.  The forward-causal aspect of the model is therefore equated with solving some dynamical equation from some initial condition, as usual.  But to add retrocausation, these models must `then' take some future input as a final condition used to dynamically solve an equation in the opposite time direction.  It is debatable whether or not this process should continue to iterate through time, may compute inconsistent values on the various iterations, or remain logically coherent without introducing a second time parameter.  

All-at-once retrocausation, on the other hand, takes a clear stand on these debatable issues.  Such models generally result from taking seriously the block-universe view of general relativity, such that any given parameter at any given location in spacetime can only have one value.  By allowing both initial and final inputs into the calculation, but only permitting one meaningful result at every intermediate spacetime location, the entire problem must be solved ``all at once'', denying any objective dynamical or causal ``flow'' from one time to another.  Such models are still accurately termed ``retrocausal'', because the future inputs are constraining past parameters, much in the way that the position of laser cavity mirrors constrain the normal modes of the cavity.\footnote{Just as we would naturally say that such normal modes were caused by all of the adjustable mirror positions, the solution of an all-at-once calculation is caused by all of the controllable boundaries on the problem, both past and future.}  If some of those boundaries take the form of future inputs (settings), then such models are strictly retrocausal, even without any objective causal ``flow'' back in time.

If the goal of a retrocausal model is to explain Bell inequality violations by correlating hidden parameters with future settings, such that $P(\lambda|\alpha,\beta)\ne P(\lambda)$, then either of these accounts of retrocausation are arguably reasonable.  While some retrocausal models of conventional Bell correlations must be solved all-at-once \cite{almada2016,wharton2018}, others seem to contain both forward- and reverse-dynamical calculations \cite{sutherland2021}.  Either way, the key is that the future settings have some adjustable constraint on the past, providing a mechanism to violate Bell Statistical Independence.

But the triangle network sheds a different light on such models.  To explain these experiments, one does not need to explain a correlation between the future settings and a past parameter, but instead an unfactorizable joint probability of the original parameters $P(\lambda_A,\lambda_B,\lambda_C)$.  The role of the future settings is conspicuously absent; no adjustable settings are even required to prove the no-go theorem of the triangle network.  Instead, the non-classical behavior seems to be governed by the global topology of the connections through space and time; it is this $\VW$ pattern which must somehow create the unusual correlations.  

Because of this feature, only an all-at-once model would seem to be available for a retrocausal account of such an experiment.  The global topology does not reside on some future setting, waiting to be used as an input in some reverse dynamical equation.  Instead, the topology is more like the connections between the nodes of an Ising model, where the full pattern is known to constrain joint probabilities  \cite{wharton2014}.  Instead of labelling the models as future-\emph{input} dependent, it might be more accurate to say that they are future-\emph{topology} dependent.  Of course, the topology of an experiment is certainly an externally-controllable input (it can be altered).  But the point remains that this input does not reside on some future state, ready to be used in reversed-dynamical equations.  When one builds the $\VW$ topology of a triangle-network experiment, one is imposing an all-at-once constraint, and it seems reasonable that only an all-at-once retrocausal model could properly accept such a constraint as an input.

\section{Summary}

The above framework makes it straightforward to analyze any entanglement experiment using the path integral (or sum-over-history) approach.  Previous work had already shown how to generate any arbitrary entangled state in a path-integral-friendly geometry \cite{tyagi2022}, but it was not obvious how to extend this to ``entangled measurements'', where multiple particles are measured in a non-separable basis.  A simple implementation of an arbitrary two-particle measurement basis was introduced in Section 2.  Combined with the previous results, any entanglement experiment can now be analyzed in a straightforward manner.

Several applications of this new toolbox were demonstrated in Section 3 -- specifically, entanglement swapping and the triangle network.  In every case, the probabilities generated by the path integral approach match the probabilities as calculated by conventional QM.  This includes joint probabilities which violate various no-go theorems.

Although the bare probabilities are the same, Section 4 detailed several topics for which this alternate technique seems to imply different conclusions about the underlying systems in these experiments.  This analysis seem likely to inform research in a wide variety of topics in quantum foundations, including: the fundamental nature of instantaneous states vs. sequential histories; foundational accounts of objective probabilities; the importance of reference frames and time-ordering; entanglement as an objective property of systems; nonlocal vs. retrocausal models; dynamical vs. all-at-once retrocausation.  One further observation was that it might be extremely useful to categorize different entanglement experiments not by their probability distributions, but rather by the topology of the spacetime connections used to implement them ($\W$ vs. $\VW$, etc.).

We expect that these results are merely the first in a series of future examinations of entanglement from this alternate perspective.  Now that all entanglement experiments can be simply analyzed in this manner, many other research avenues appear to be opened.  It may turn out that the path integral provides a no-more-satisfactory account of these experiments than does conventional QM, but at this point there seems little reason to prefer one method over the other.  By analyzing their differences, as well as their points of agreement, we hope that the resulting implications will motivate further novel accounts of entanglement experiments.

\section*{Acknowledgements}

The authors would like to thank N. Argaman for bringing the intriguing features of the triangle network to our attention (and other useful advice), H. Price for encouraging us to think more carefully about entanglement swapping, E. Adlam for very useful insights, and of course N. Tyagi for her work on which this paper was based.

\bibliographystyle{unsrt}
\bibliography{References.bib}

\end{document}